\documentstyle[12pt,epsfig]{article}
\begin{document}
\begin{titlepage}

\large
\centerline {\bf Physics Prospects at BTeV}
\normalsize
 
\vskip 2.0cm
\centerline {Penelope Kasper\footnote{Email penny@fnal.gov}}
\vspace{3.0mm}
\centerline {\it Fermilab}
\centerline {\it Batavia, IL 60510, USA}
\vskip 1.5cm
\centerline {For the BTeV Collaboration}
\vskip 2.0cm
\centerline {\bf Abstract}
\vskip 1.0cm 
  BTeV is a proposed forward collider program at the Fermilab Tevatron
  dedicated to precision studies of CP violation, mixing and rare decays 
   of beauty and charm hadrons.

\vfill
\end{titlepage}

\section{Introduction}
  Although most data on $B$ physics has been collected at $e^+e^-$ colliders
  there is much interest in doing $B$ physics at hadron colliders because of
  the large production cross-section. 
   At the Fermilab Tevatron, CDF and D0 have produced many 
  interesting B physics results proving the feasibility of doing $B$ physics
  at a hadron collider.
  In the next few years
  current experiments CLEO, CDF and D0 will begin new runs with upgraded detectors
  and several new detectors (HERA-B, BELLE and BaBar) will begin data taking.
  However there are many channels that are beyond the limits of sensitivity 
  of these experiments. In addition, the $e^+ e^-$ machines operating 
  at the $\Upsilon(4S)$ cannot produce $B_s$, $B_c$ or $\Lambda_b$.
  
    BTeV is intended to be a next generation  high statistics
   heavy quark experiment. The design of the proposed forward collider detector is optimized
   for precision studies of beauty and charm decays and is designed to fit
   in the
   C0 interaction region of the Tevatron. 
   An Expression of Interest (EOI) was submitted to the Fermilab PAC in May 1997 \cite{EOI}.
   In January 1998 BTeV was formally approved as an R\&D project and a
   contract was awarded for the
   construction of an appropriate collision hall in the C0 interaction region.

\section{Physics Goals}
  The main physics goals of BTeV are the precise measurement of Standard Model parameters
  including CKM matrix elements and decay constants, and the search for new phenomena beyond 
  the Standard Model.
     We assume that the angle $\beta$ of the unitarity triangle will already have been
    measured, but we will be able to reduce the error. 
    In particular, we will measure the asymmetry in $B^0 \rightarrow \pi^- \pi^+$ and
     the angle $\gamma$ of the unitarity triangle.
    Time dependent studies of mixing in $B_s$ decays 
   will  lead  to the 
   measurement of $x_s$ ($\Delta{m}/\Gamma$), and we will also measure 
   the width difference $\Delta\Gamma$
   between the two CP eigenstates.
    In the Standard Model CP violating effects are small in the charm sector so this
    provides an excellent opportunity to search for non Standard Model effects.

\section{Detector Design}
  The Fermilab Tevatron is the highest energy particle accelerator in operation.
  The new Main Injector
  will increase the luminosity to $2\times 10^{32}$cm$^{-2}$s$^{-1}$.
  The simulations for BTeV assume a luminosity of 
  $5\times 10^{31}$cm$^{-2}$s$^{-1}$
   which results
  in a mean number of interactions per crossing of 0.5.
  At a collider energy of 2 TeV the $b\bar{b}$ production cross section is 
  $\approx 100\mu$b.
  The Tevatron operating at a luminosity of  $5\times 10^{31}$cm$^{-2}$s$^{-1}$
   will yield
  $5 \times 10^{10}$ $b\bar{b}$ pairs per year.
     Eventually we intend to run with a luminosity of $2\times 10^{32}$cm$^{-2}$s$^{-1}$
  under luminosity leveled conditions \cite{GJ} which gives an effective factor of 8
  increase in the number of events. 

\begin{figure}
 \centerline{ \epsfxsize=6.0in \epsffile{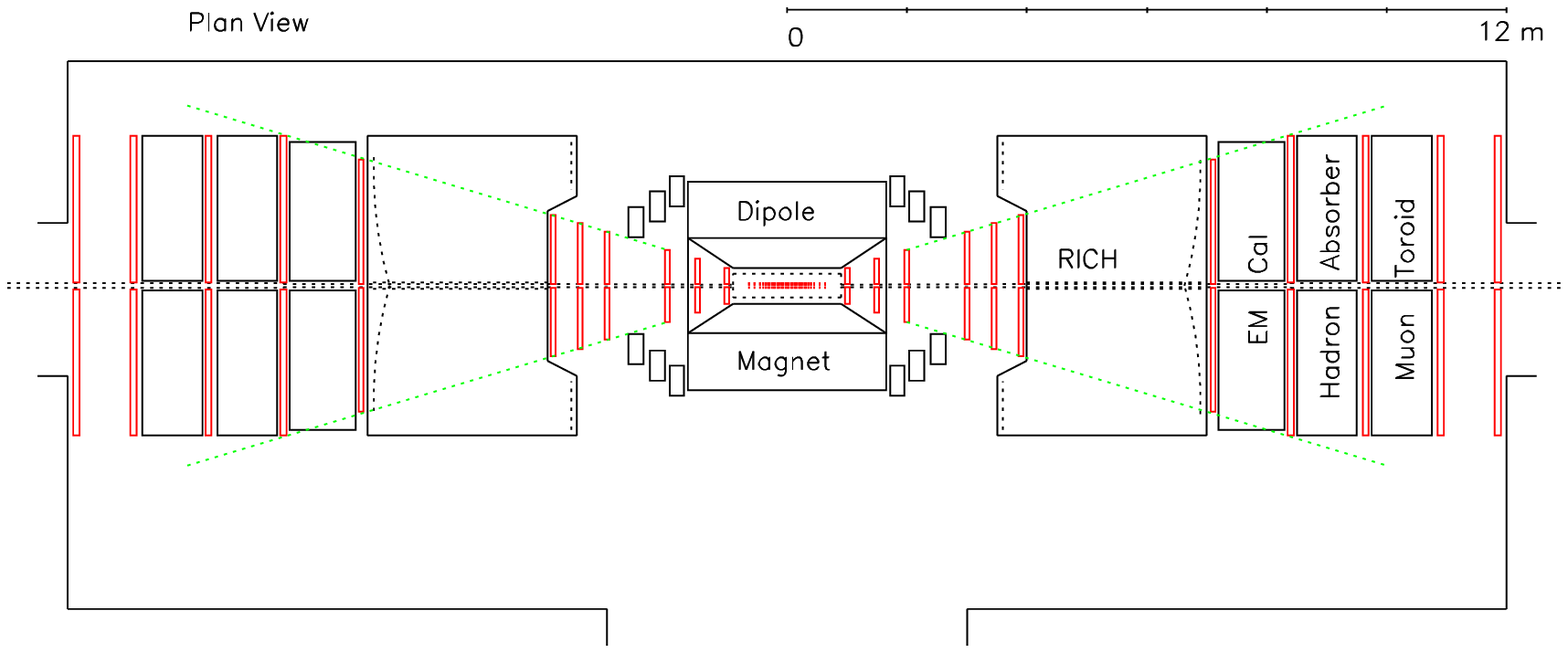} }
\caption{Schematic Layout of the BTeV detector}
\label{fig:spect}
\end{figure}

   The proposed reference detector 
  is a 2-arm forward collider detector ($1.5<|\eta|<4.5$)(Fig.~\ref{fig:spect})
   There are several advantages of having a forward detector: (i) the higher momentum forward
  $B$'s have a greater vertex separation - this enables us to have a Level I 
   vertex trigger, helps to reduce background and
   improves time resolution,
  (ii)  there is room for charged particle identification, and
  (iii) the strong correlation between the $b$ and $\bar{b}$ production angles
  in the forward direction means that a significant fraction of events will have both 
  $b$ and $\bar{b}$ in the detector.
  
\begin{figure}
 \centerline{ \epsfxsize=3.0in \epsffile{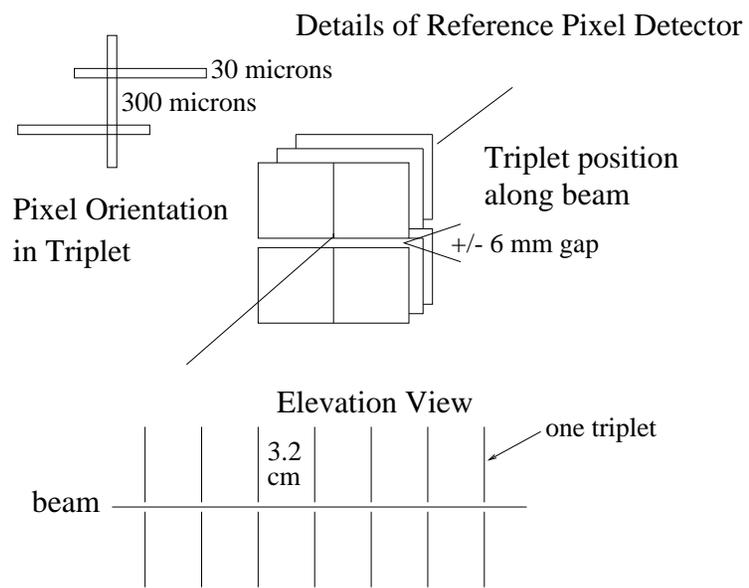} }
\caption{Schematic of the BTeV vertex detector}
\label{fig:pixels}
\end{figure}

   The interaction region has a length of $\sigma_z =30$ cm and we have a
    pixel vertex detector inside a central dipole field which surrounds the interaction 
   region. This enables a rough momentum measurement on the tracks in the
   Level I vertex trigger.
       The layout of the pixel planes is shown in Fig.~\ref{fig:pixels}.
    Triplets of silicon pixel planes will be placed inside the beam pipe, as close as possible to 
    the beam to ensure
    a large acceptance in the forward direction. The planes are retracted while the
    collider is being filled and then closed down to leave a gap of $\pm6$ mm.
    New simulations indicate a significant improvement in $x_s$ reach if a
    detector with a square
    hole, $1.2$cm $\times$ $1.2$cm, centered on the beamline, is used.

   Silicon pixel detectors rather than silicon strips are  chosen because of their 
   low occupancy, superior
   signal to noise, and intrinsically better pattern recognition which enhances the
   ability to do real-time tracking that is needed for the Level I vertex trigger.
   The reference detector has pixels $30\times300\mu$m but we are studying
   how much bigger the pixels can be made without degrading the performance.
   A more detailed account of the continuing pixel R\&D effort is given
   in these proceedings \cite{SKWAN}. 
   Downstream tracking is necessary
   to  provide better momentum resolution and to reconstruct $K_s$'s which decay
   outside the pixel detector. We are considering different technologies including
   straw tubes.
   
   Charged particle identification is essential to distinguish between final states
   which overlap in mass,
    for example $B_d^0 \rightarrow \pi^+ \pi^-, B_d^0 \rightarrow K^{\pm} \pi^{\mp}$ and
   $B_s^0 \rightarrow K^+ K^-$. In addition, many studies of neutral $B$ mesons require
   flavor tagging. Our studies show that kaon tagging is a very effective method. 
   The proposed RICH is described in these proceedings \cite{TOMASZ}.
    Design studies for a muon detector and electromagnetic calorimeter are also in progress.

\begin{figure}
 \centerline{ \epsfxsize=3.0in \epsffile{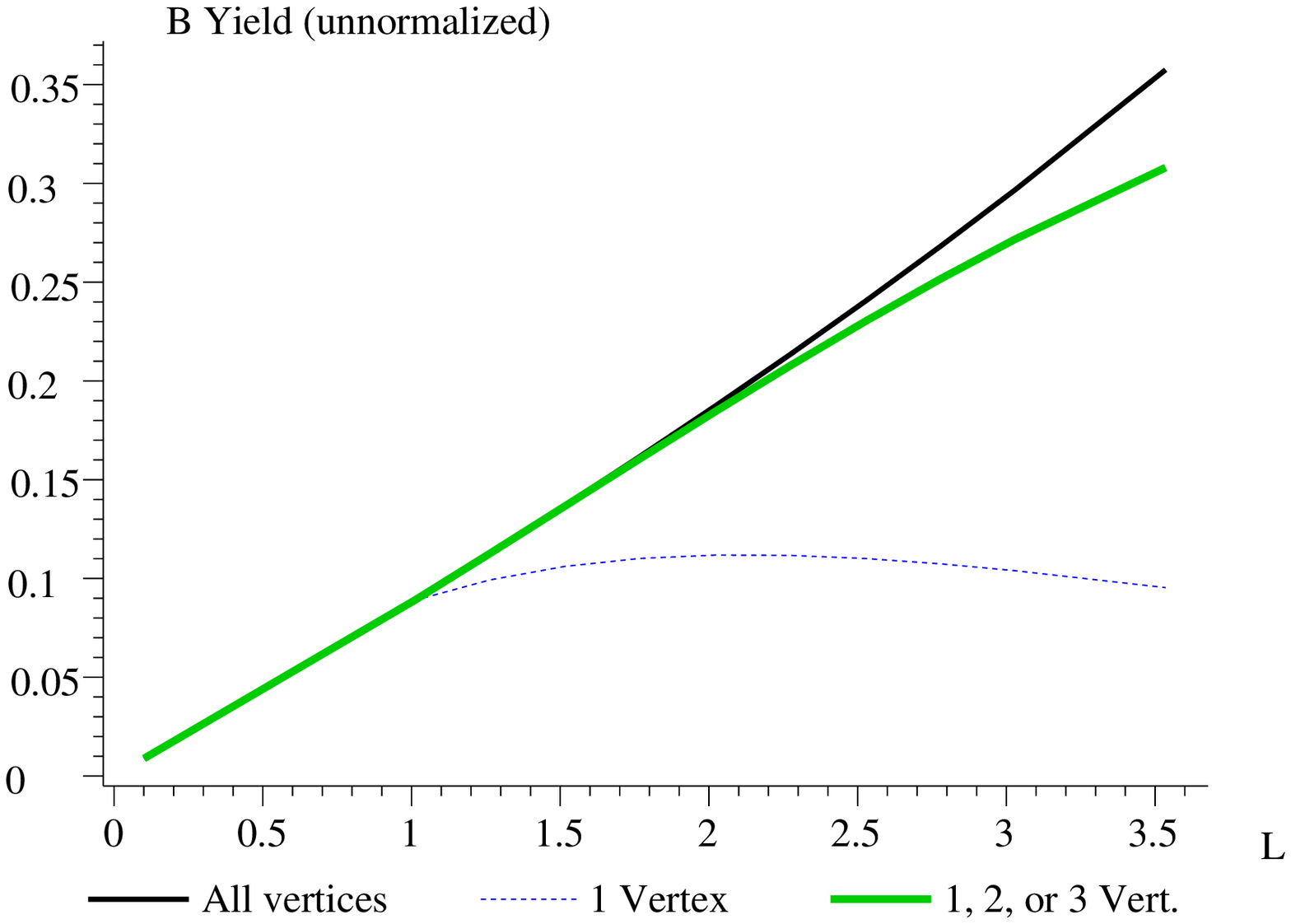} }
 \centerline{ \epsfxsize=3.0in \epsffile{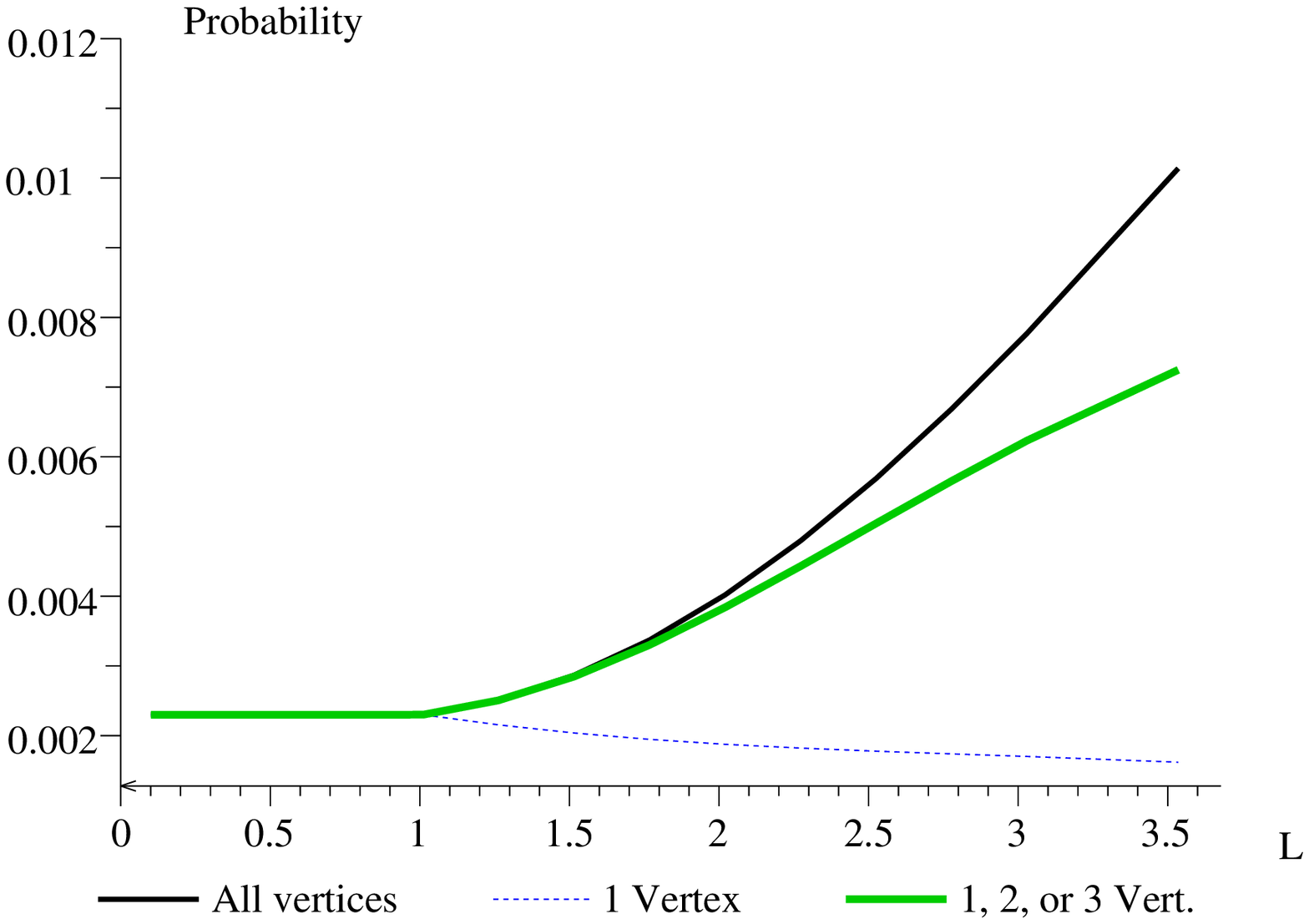} }
\caption{Top curve: The  yield  for events with
$B^{o}\rightarrow \pi^{+}\pi^{-}$. Since the curves are not 
absolutely normalized, their important features are their shapes.
Bottom curve: The absolute probability  for triggering on crossings
with inelastic interactions.  For both plots, the lower axis is
the  luminosity in units of $10^{32}\,{\rm cm}^{-2}\,{\rm s}^{-1}$.
The solid line is for events with $\ge 1$ reconstructed
vertex; the dotted line, for events with  one and only one reconstructed 
vertex; and the thick gray line for events 1, 2, or 3 reconstructed vertices.} 
\label{fig:trig}
\end{figure}

   An essential component of BTeV is the Level I vertex trigger.
   The requirements for the trigger are that it must be capable of reducing 
   the event rate by
    a factor of more than 100 while maintaining a high efficiency for heavy quark
    events which pass the offline analysis.
       The trigger is based on having a number of tracks with significant impact parameter
    with respect to the primary vertex. The reference trigger scheme was developed by
    the University of Pennsylvania \cite{TRIG} and is currently being refined.
    The effects of multiple interactions per bunch crossing and secondary
    interactions in the vertex detector are being studied.
      Initial results on the performance of the baseline trigger in the presence of
      multiple interactions are shown in Fig~\ref{fig:trig}. The yield for $B^0 \rightarrow \pi^+ \pi^-$
      is linear in luminosity while the probability for triggering on a minimum bias event
      is 0.23\% up to a luminosity of $10^{32}$cm$^{-2}$s$^{-1}$ and then rises to 0.40\% at
      a luminosity of $2\times10^{32}$cm$^{-2}$s$^{-1}$ \cite{multin}.

\section{Simulations}
    Detailed simulations of several physics channels have been carried out to explore
    the physics reach of the baseline BTeV detector as described above. The results 
    are based on an average
    luminosity of $5\times 10^{31}$cm$^{-2}$s$^{-1}$ for $10^7$ s
     and a $b\bar{b}$ cross-section of 100$\mu$b. 
    However the integrated life of the experiment will be 
     much longer than $10^7$ s, and the detector
     will be able to run with a factor of 4 higher luminosity. We also plan to run under luminosity
     levelled conditions which results in an effective factor of 2 increase in luminosity.

     Events are generated using Pythia 5.7 and Jetset 7.4 \cite{PYTHIA}. The heavy quark decays are
     then modeled through the CLEO decay Monte Carlo QQ.
     The detector is simulated using MCFAST v$2\_6$ \cite{MCF}, a fast Monte Carlo package developed
     by the Fermilab Simulation Group for detector design studies. 
       	  Particle trajectories are traced though simple geometric shapes 
	  then tracks are  parametrized using a Kalman Filter technique.
	 A covariance matrix for each track is found  taking account of
	 multiple scattering and detector resolution.
          The covariance matrix is used to smear the original track parameters then
     the smeared track parameters and covariance matrix are stored as a reconstructed track.
      Hit generation is used for trigger studies.
      
\subsection{Tagging}
  In order to study neutral $B$ meson oscillations it is necessary to know
  the flavor of the $b$ quark ($b$ or $\bar{b}$) both at production and decay.
   Measurements of CP asymmetry in modes such as $B_d \rightarrow \pi^+ \pi^-$
   and $B_d \rightarrow \psi K_s$ also require the determination of the flavor of
   the $b$ quark at the production point. This is known as flavor tagging.
    ``Away-side'' tagging methods rely on the determination of the flavor of the
    other $b$ quark in the event {\it{eg.}} from the charge of the lepton in semileptonic
    decay or from the charge of the kaon produced in the $b\rightarrow c \rightarrow s$ cascade.
    Simulations indicate that we can expect
    an  effective tagging efficiency $\epsilon{D^2}$ of $\approx$1.5\% for
    muon tagging and $\approx$5\% for kaon tagging. Other tagging methods  are being
    studied.

\subsection{Measurement of the $B_s$ mixing parameter $x_s$ }
BTeV will have excellent resolution for secondary vertices which translates
into good proper time resolution.   This makes it an ideal detector for
studying $B_s$ mixing since $x_s$ is expected to be large.   Current limits
on $B_s$ mixing from LEP indicate that $x_s>15$ \cite{LEP-B}. Studies
of two different
 decay modes of the $B_s$  have been made in order
to understand the sensitivity to $B_s$ oscillations. 
  The decay mode 
  $B_s \rightarrow \psi \bar{K}^{*0}, \psi \rightarrow \mu^+ \mu^-$.
  was chosen
  because it had single 4-prong vertex that could be triggered both with a muon trigger
  and a secondary vertex trigger.
  From simulations we expect 220 events (reconstructed, triggered and tagged)
  per $10^7$ s in this mode with S/B=3 and a time resolution 
  of 45 fs \cite{EOI}.
  (The yield would increase by at least 50\% if the decay mode 
  $\psi \rightarrow e^+ e^-$ is also used.) 
  Later studies were done with the decay mode
   $B_s \rightarrow D_s \pi, D_s \rightarrow \phi \pi, \phi \rightarrow K^+ K^-$ which has a much
  larger branching fraction. In this mode we expect 2400 events per $10^7$ s with S/B=3
  and time resolution of 52 fs \cite{PAC}.
  
\begin{figure}
 \centerline{ \epsfxsize=3.0in \epsffile{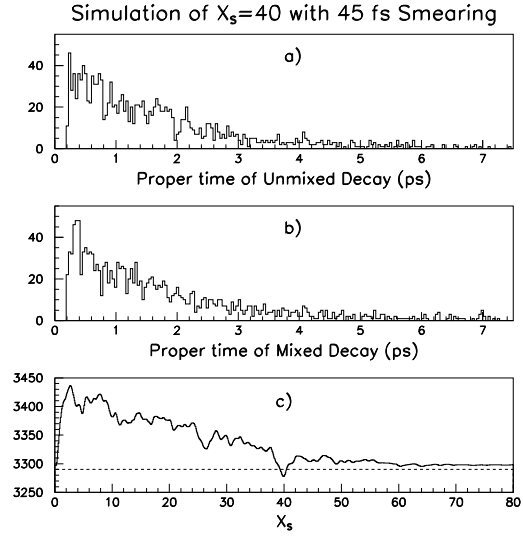} }
\caption{Proper lifetime plots of a) unmixed and b) mixed decays for the 
decay mode $B_s \rightarrow D_s \pi$ for
one year of running.  
 c) the corresponding negative log likelihood
 as a function of $x_s$. }
\label{fig:time}
\end{figure}

\begin{figure}
 \centerline{ \epsfxsize=3.0in \epsffile{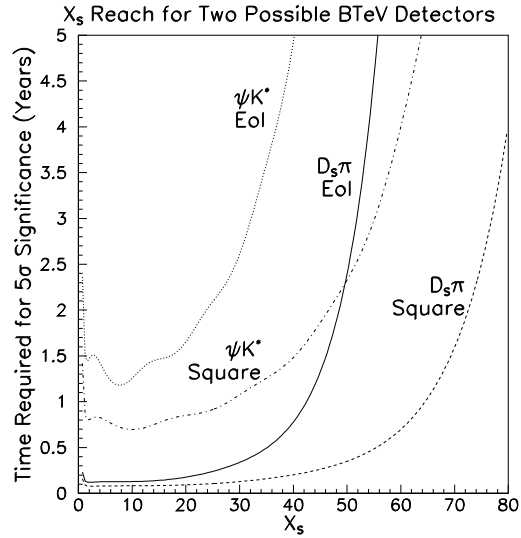} }
\caption{$x_s$ reach of the BTeV detector }
\label{fig:xs_reach}
\end{figure}

  Once the number of events, background, time resolution, tagging efficiency and
  mistag fraction were determined, a 
  mini-Monte Carlo was used to generate the expected proper time distributions
  for a particular value of $x_s$.
  Fig.~\ref{fig:time}(a) and (b) shows the mixed and unmixed time distributions for the 
  $D_s \pi$ mode generated with $x_s$=40.
  Fig.~\ref{fig:time}(c) shows the negative log likelihood function computed from these events. 
  It can be
  seen that the fit picks out the correct solution at $x_s$=40. 
  The mini-Monte Carlo also shows how the limiting $x_s$ sensitivity of the experiment
  is approached. As the number of events is reduced, the negative log likelihood 
  function becomes
  more and more ragged and the depth of the secondary minima approach that of the
  true minimum.
  We define a significant
  observation as one in which the deepest minimum is deeper than the next deepest
  by at least $5\sigma$. This is shown by the dashed line. When a solution is
  found the error is $\approx 0.1$.
   Fig.~\ref{fig:xs_reach} shows the number of years needed to obtain a significant
   measurement as a function of $x_s$. This figure also shows the improvement
   in $x_s$ reach obtained with the square hole pixel layout.

\subsection{ CP Violation in $B^0 \rightarrow \pi^+ \pi^-$ }  
\begin{figure}
 \centerline{ \epsfxsize=3.0in \epsffile{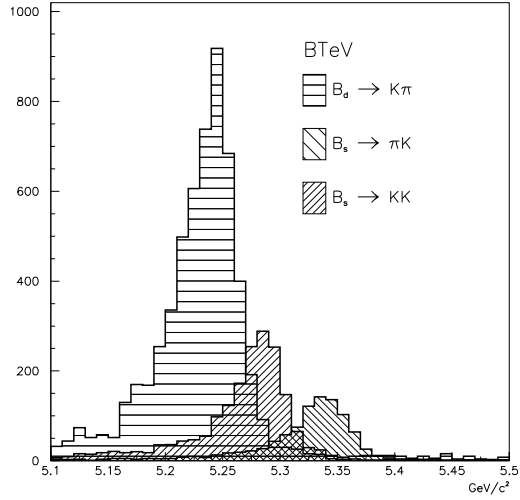} }
 \centerline{ \epsfxsize=3.0in \epsffile{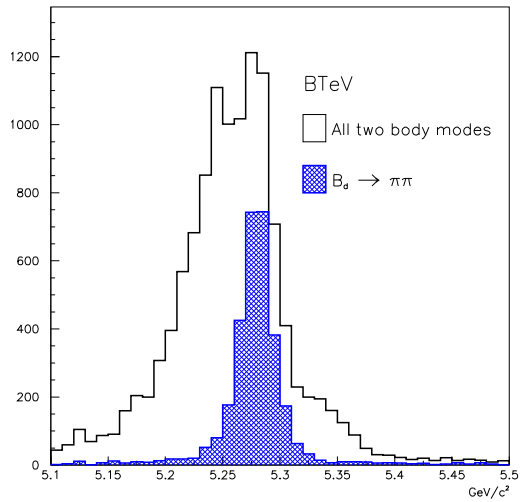} }
 \caption{ Two body mass plots without particle identification a) including
contributions from  $B_d \rightarrow K^+\pi^- $,
$B_s \rightarrow \pi^+K^- $, $B_s \rightarrow K^+K^- $, b) $B_d \rightarrow
\pi^+\pi^-$ and a sum of all two body decay modes.  All particles are
assumed to be pions.  }
 \label{fig:pipi}
\end{figure}

   Measurement of the CP asymmetry in the decay $B^0 \rightarrow \pi^+ \pi^-$
   leads to information about the angle $\alpha$ of the unitarity triangle.
   This decay mode is also an example of a low multiplicity final state which provides
   a critical test of the Level I trigger algorithm.
   The large combinatoric background expected for this mode can be reduced by requiring that the
   secondary vertex be well separated from the primary and that the reconstructed
   $B^0$ point back to the primary vertex. Another large source of background
   comes from other 2-prong $B$ decays - $B_d \rightarrow K^+ \pi^-$, $B_s \rightarrow K^+ K^-$,
   and $B_d \rightarrow K^+ \pi^-$. This background can be reduced below the level of 
   combinatoric background with good kaon identification in the RICH.
   Fig.~\ref{fig:pipi}(a) shows the contribution to the two pion mass plot from each of these
   decay modes. Fig.~\ref{fig:pipi}(b) shows the signal compared to the sum of all two body
   modes.
   
     Simulations result in a combined geometric acceptance and reconstruction
     efficiency of 9.0\% and a trigger efficiency for accepted events of 72\% \cite{EOI}.
     Assuming a branching ratio of $0.75 \times 10^{-5}$ then
     we expect 16,000 triggered and reconstructed events per year before tagging.
     Background studies of $5 \times 10^6$ generic $B$ events lead to an
      estimate of S/B=0.9.
     If we assume an effective tagging efficiency $\epsilon{D^2} = 10\%$ then
     the statistical error in the time-integrated CP asymmetry is  
     $\sigma_{A_{CP}}=\pm0.04$.
     If only tree level  diagrams contribute to this decay then $\alpha$ 
     is related to the asymmetry
     by 
       \[ A_{CP} =  \sin{2\alpha}\frac{x_d}{1+x_d^2},    \] 
      however it is expected that contributions from penguin diagrams will not be
      negligible and other decay modes will need to be studied to determine
      the penguin effect.

\subsection{ Measurement of $\gamma$ }
  The angle $\gamma$ of the unitarity triangle is the most difficult to measure
  experimentally and all methods result in discrete ambiguities. 
   Using
    several different methods will help resolve these ambiguities.
  We have studied
  three methods of measuring $\gamma$.
  The first method uses the decays $B_s \rightarrow D_s^{\pm} K^{\mp}$ where a 
  time-dependent CP violation can result from
  the interference between the direct decay and the mixing induced decay \cite{Aleks}.
   Consider the following time-dependent rates : 
\[  
  {\Gamma}(B_s\rightarrow f) = |M|^2 e^{-{t}}
  \{ \cos^2(xt/2) + \rho^2\sin^2(xt/2) - \rho{\sin(\phi+\delta)}\sin(xt) \}  
\]  
\[  \Gamma(\bar{B_s}\rightarrow \bar{f}) = |M|^2 e^{-{t}}
  \{ \cos^2(xt/2) + \rho^2\sin^2(xt/2) + \rho{\sin(\phi-\delta)}\sin(xt) \}  \]
\[  \Gamma(B_s\rightarrow \bar{f}) = |M|^2 e^{-{t}}
  \{ \rho^2\cos^2(xt/2) + \sin^2(xt/2) - \rho{\sin(\phi-\delta)}\sin(xt) \}  \]
\[  \Gamma(\bar{B_s}\rightarrow f) = |M|^2 e^{-{t}}
  \{ \rho^2\cos^2(xt/2) + \sin^2(xt/2) + \rho{\sin(\phi+\delta)}\sin(xt) \},  \]
where  $M = \langle~f|B\rangle$, 
        $\rho = \frac{\langle~f|\bar{B}\rangle}{\langle~f|{B}\rangle}$,
        $\phi$ is the weak phase between the 2 amplitudes and
        $\delta$ is the strong phase between the 2 amplitudes.
   The three parameters $\rho$, $\sin(\phi+\delta), \sin(\phi-\delta)$
   can be extracted 
   from a time-dependent study if $\rho=O(1)$. 
        
   In the case of $B_s$ decays  
  where $f=D_s^+ K^-$ and  $\bar{f}= D_s^- K^+$,
    the weak phase is $\gamma$.
   The decay mode $B_s \rightarrow D_s K$, $D_s \rightarrow \phi \pi$, $\phi \rightarrow K^+ K^-$ 
   was simulated and resulted in
   a combined geometric acceptance
   and reconstruction efficiency of 4\% with S/B=10 \cite{Bsdsk}.
    The trigger efficiency is 70\%.
   Using the branching fractions from Aleksan \cite{Aleks} and
   assuming a tagging efficiency $\epsilon = 15\%$ we expect 350 tagged events per year.
   We have a total of 800 events, with S/B= 10, if we also use the $K^* K$ decay mode of the $D_s$. 
   
\begin{figure}
\centerline{ \epsfxsize=5.0in \epsffile{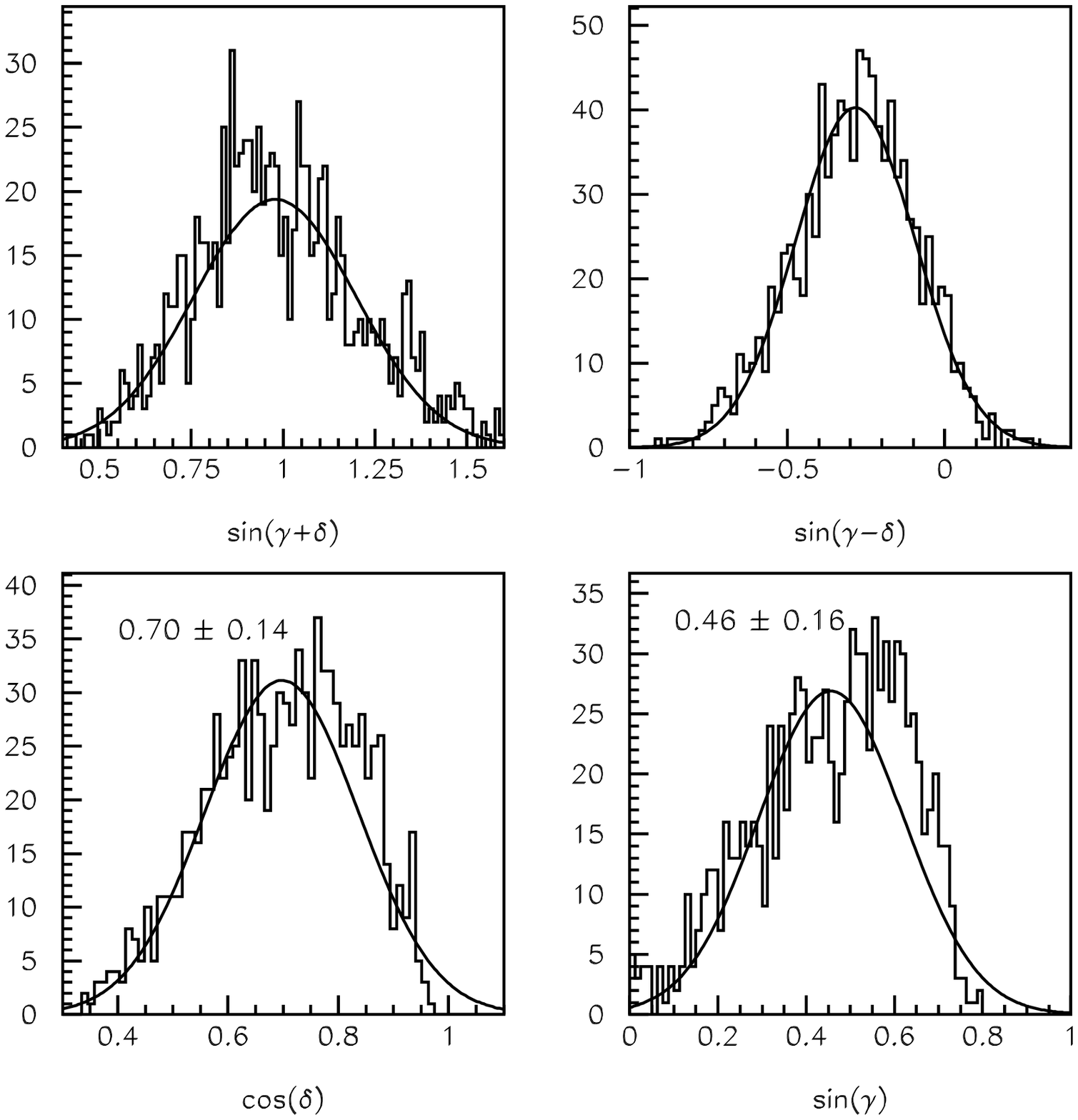} }
\caption{$B_s \rightarrow D_s K$, distributions of fitted parameters, for a signal
of 1000 events and input values
 $\sin(\gamma+\delta)= .97$,
 $\sin(\gamma-\delta)=-.28$, $\sin\gamma=.5$, $\cos\delta=.7$  }
\label{fig:gamma}
\end{figure}

   Using the measured values of S/B and time resolution, a mistag rate of 25\%,
   and $x_s$=40,
   a mini-Monte Carlo was used to generate the expected proper time distributions
   for various sets of input parameters $\rho$, $\sin(\gamma+\delta), \sin(\gamma-\delta)$.
   A maximum likelihood fit was then used to extract fitted values of the parameters.
   Fig.\ref{fig:gamma} shows the distributions of the parameters for a signal of 1000 events
   with input values $\rho$=0.5, $\sin\gamma=0.5$ and $\cos\delta=0.7$.
   Assuming that $\sin\gamma >0$ then $\sin\gamma$ can be determined up to a two-fold
   ambiguity, hence $\gamma$ up to a four-fold ambiguity.
   
     The second method
    described by Gronau and Rosner \cite{GR} and Fleischer and Mannel \cite{FM}
    uses $B^0 \rightarrow K^+ \pi^-$ and $B^+ \rightarrow K^0 \pi^+$ decays. It is particularly
    promising as it may complement other methods by excluding some of the
    region around $\gamma=\pi/2$. 
     We expect to reconstruct 3600 $B^{\pm}\rightarrow K_s \pi^{\pm}$ with 
     S/B=0.5
     and 36000 $B^0/\bar{B}^0 \rightarrow K^{\pm} \pi^{\mp}$ with
     S/B=3. 
     Gronau and Rosner estimate a measurement of $\gamma$ to $10^{\circ}$
     with 2400 events in each channel \cite{Ros}, however there has been much theoretical
     discussion about the effects of isospin conservation and rescattering which casts
     doubt on this method
     \cite{Gerard}\cite{FK}\cite{MNeub}\cite{Atsoni}.

    Another method  for extracting $\gamma$ has been proposed by
    Atwood, Duneitz and Soni \cite{ADS}.    
   A large CP asymmetry can result from the interference of the
   decays $B^- \rightarrow K^- D^0, D^0 \rightarrow f$ and 
   $B^- \rightarrow K^- \bar{D}^0, \bar{D}^0 \rightarrow f$, where $f$ is
   a doubly Cabibbo suppressed decay of the $D^0$  
   (for example $f= K^+\pi^-, K\pi\pi$, etc.) 
   Since $B^- \rightarrow K^- \bar{D}^0$ is color-suppressed and $B^- \rightarrow K^- D^0$
   is color-allowed, the overall amplitudes for the two decays are
   expected to be approximately equal in magnitude.  The weak phase
   difference between them is $\gamma$. To observe a CP asymmetry there
   must also be a non-zero strong phase between the two amplitudes.
     It is necessary to measure the branching ratio 
     ${\cal B}(B^- \rightarrow K^- f)$ for at least 2 different states $f$ in order
     to determine $\gamma$ up to discrete ambiguities. 
     We have examined the decay modes $B^- \rightarrow K^- [K^+ \pi^-]$
     and $B^- \rightarrow K^- [K^+ 3\pi]$. The combined geometric acceptance
     and reconstruction efficiency was found to be 4\% for the $K \pi$ mode and
     3.4\% for $K 3\pi$ with a signal to background of about 1. 
     The trigger efficiency is approximately 70\% for both modes.
     The expected number of $B^{\pm}$ events in $10^7$~s is 190 in the $K \pi$
     mode and 320 in the $K 3\pi$ mode.
      With this number of events we expect to be able to measure $\gamma$ 
     (up to
      discrete ambiguities)  with a statistical error of about ${\pm}20^{\circ}$
       in one year of running
   at ${\cal L}=5\times 10^{31} {\rm cm}^{-2}{\rm s}^{-1}$. 
    The overall sensitivity
   depends on the actual values of $\gamma$ and the strong phases.

\section{Conclusions}
    Hadron collider experiments provide the best environment to obtain
    the high statistics needed for precision measurents of CP Violation
    and $B_s$ mixing. 
      The forward detector  of BTeV has been designed to
      fit in the new C0 interaction region at the Tevatron and 
      incorporates a silicon pixel
    vertex detector,  downstream tracking and charged paricle
    identification. The vertex detector enables  Level I vertex triggering
    and excellent time resolution. 
       The strength
      of the good time resolution is reflected by the fact that
       $x_s$ can be measured up to
    a value of 60 in one year of  running at a luminosity of 
    $5\times10^{31}$cm$^{-2}$s$^{-1}$.
    Eventually we plan to run with a luminosity of $2\times10^{32}$cm$^{-2}$s$^{-1}$
    under luminosity leveled conditions which results in a factor of 8 increase in
    the number of events per year.
      A summary of the physics reach is shown in Table~\ref{tab:phys}. 

\begin{table}
\caption{BTeV Physics Reach}
\begin{center}
\label{tab:phys}
\begin{tabular}[t]{|c|c|c|}
\hline
Measurement & Accuracy in $10^7$ s & Accuracy in $10^7$ s \@ \\
            & \@ ${\cal L}=5\times 10^{31}$ & ${\cal L}=2\times 10^{32}$, $\cal L$ leveled \\
\hline
$x_s$ (square hole)            & up to 64   & up to 80 \& beyond \\

$A_{CP}(B^0\rightarrow\pi^+\pi^-)$    & $\pm 0.036$ & $\pm 0.013$ \\

$\sin\gamma$ using $D_s K^-$  & $\pm{0.14}^{\dag}$ & $\pm{0.05}^{\dag}$ \\

$\gamma$ using $D^0 K^-$      &  $\pm 20^{\circ\ddag}$ & $\pm 7^{\circ\ddag}$ \\

${\cal B}(B^- \rightarrow K^-\mu^+\mu^-) $  & $4\sigma$ at ${\cal B}$ of $1.4\times10^{-7}$ & 
$4\sigma$ at ${\cal B}$ of $5.4\times10^{-8}$ \\
\hline
\multicolumn{3}{l}{\dag Assumes $\rho$=0.7, $\cos\delta$=0.7, $\sin\gamma$=0.5, $x_s$=20} \\
\multicolumn{3}{l}{\ddag For most values of strong phases and $\gamma$} \\
\end{tabular}
\end{center}
\end{table}

\section*{Acknowledgements}
  The author would like to thank the members of the BTeV collaboration who
   have contributed to this work;
  in particular Rob Kutschke, Mike Procario, Paul Lebrun, 
  Tomasz Skwarnicki, Ramazan Isik and Kevin Sterner whose results I have
  quoted, and Patty McBride and Sheldon Stone for help in preparing this
  paper.

\clearpage

\clearpage
     
  
\begin{thebibliography} {99}
\bibitem{EOI} BTeV EOI (May 1997), http://www-btev.fnal.gov/btev.html

\bibitem{GJ}G. Jackson, ``A Dedicated Hadronic B-factory'', these proceedings

\bibitem{SKWAN}S. Kwan, ``BTeV Pixel Development and Triggering'', these proceedings

\bibitem{TOMASZ}T. Skwarnicki, ``BTeV RICH'', these proceedings

\bibitem{TRIG}R. Isik, W. Selove and K. Sterner,
``Monte Carlo Results for a Secondary-vertex Trigger with On-line
 Tracking'',  Univ. of Penn. preprint UPR-234E(1996)

\bibitem{multin}P. Lebrun, BTeV-int-97/17 and BTeV-int-98/2 

\bibitem{PYTHIA}H. U. Bengtsson and T. Sjostsrand, Comput. Phys. Commun, {\bf{46}}, 43(1987)

\bibitem{MCF}
 P. Avery et al., ``MCFast: A Fast Simulation Package for Detector
Design Studies'' in the Proceedings of the International Conference on Computing
in High Energy Physics, Berlin, (1997) and 
http://fnpspa.fnal.gov/mcfast/mcfast\_docs.html

\bibitem{LEP-B} V. Andreev {\it et al.} (The LEP $B$ Oscillations Working Group), ``Combined Results
on $B^0$ Oscillations: Update for the Summer 1997 Conferences,'' LEP-BOSC 97/2, August
18, 1997.

\bibitem{pipi}M. Procario, BTeV-int-97/10

\bibitem{PAC} BTeV response to PAC (December 1997)

\bibitem{Bsdsk}P. A. Kasper, BTeV-int-97/15

\bibitem{Aleks}R. Aleksan, I. Dunietz, B. Kayser,
``Determining the CP-violating phase $\gamma$'',
Z. Phys C {\bf 54}, 653-659 (1992)

\bibitem{GR} M. Gronau and J. Rosner, CALT-68-2142, hep-ph/9711246 

\bibitem{FM} R. Fleischer and T. Mannel, hep-ph/9704423

\bibitem{Ros} J. Rosner, private communication

\bibitem{Gerard} J.-M. Gerard and J. Weyers, ``Isospin amplitudes and CP violation
 in $(B \to K\pi)$ decays," hep-ph/9711469 (1997).

\bibitem{FK} A. Falk, A. Kagan, Y. Nir and A. Petrov, JHU-TIPAC-97018
(December 1997).

\bibitem{MNeub}M. Neubert, ``Rescattering Effects, Isospin Relations and Electroweak
 Penguins in $B \to \pi K$ Decays,"  hep-ph/9712224 (1997).

\bibitem{Atsoni} D. Atwood and A. Soni, ``The Possibility of Large Direct CP Violation
 in $B \to K \pi$-Like Modes Due to Long Distance Rescattering Effects and 
 Implications for the Angle $\gamma$," hep-ph/9712287 (1997).

\bibitem{ADS} D. Atwood, I. Dunietz and A. Soni, PRL 78, 3257(1997).

\end{thebibliography}
 \end{document}